\documentclass[12pt]{nature_weaver_preprint}

\usepackage{
aasdefs,
deluxetable,
graphicx,
color}

\newcommand{\bc}{\begin{center}}
\newcommand{\ec}{\end{center}}

\newcommand{\ie}{\textit{i.e.,\ }}
\newcommand{\eg}{\textit{e.g.,\ }}

\newcommand{\etal}{\textit{et al.\ }}

\newcommand{\hst}{\textit{HST\ }}

\begin{document}


\noindent
{\large\bf The Discovery of Two New Satellites of Pluto}\\
 
\noindent
H. A. Weaver\footnote{The Johns Hopkins University 
Applied Physics Laboratory,
Space Department, 11100 Johns Hopkins Road, Laurel, MD 20723-6099.
$^{2}$Southwest Research Institute, Space Science and Engineering Division,
1050 Walnut Street, Suite 400, Boulder, CO 80302.
$^{3}$Space Telescope Science Institute, 3700 San Martin Drive,
Baltimore, MD 21218.
$^{4}$Southwest Research Institute, Department of Space Studies,
1050 Walnut Street, Suite 400, Boulder, CO 80302.
$^{5}$Lowell Observatory, 1400 W. Mars Hill Road, Flagstaff, AZ 86001.},
S. A. Stern$^{2}$,
M. J. Mutchler$^{3}$,
A. J. Steffl$^{4}$,
M. W. Buie$^{5}$,
W. J. Merline$^{4}$,\\
J. R. Spencer$^{4}$,
E. F. Young$^{4}$,
\&  L. A. Young$^{4}$\\

\noindent
{\it Preprint version of a paper accepted for publication in the journal Nature}\\




\begin{bfseries}
\begin{boldmath}

\noindent
Pluto's first known moon, Charon, was discovered in 1978\cite{Christy:1978} and has a
diameter about half that of Pluto\cite{Buie:1992,Young:1994a,Sicardy:2005}, 
which makes it larger relative to its primary
than any other moon in the Solar System.
Previous searches for other satellites around Pluto have been 
unsuccessful\cite{Stern:1991, Stern:1994, Stern:2003},
but they were not sensitive to objects $\la$150~km in diameter
and there are no fundamental reasons why Pluto should
not have more satellites\cite{Stern:1994}.
Here we report the discovery of two additional moons around Pluto, provisionally designated \mbox{S/2005 P1} (hereafter P1) and \mbox{S/2005 P2} (hereafter P2),
which makes Pluto the first Kuiper belt object (KBO) known to have multiple satellites.
These new satellites are much smaller than Charon (diameter $\sim$1200~km),
with P1 ranging in diameter from 60--165~km depending on the surface reflectivity,
and P2 about 20\% smaller than P1.
Although definitive orbits cannot be derived,
both new satellites appear to be moving in circular orbits
in the same orbital plane as Charon, with orbital periods of $\sim$38~days (P1)
and $\sim$25~days (P2).
The implications of the discovery of P1 and P2 for the origin and evolution 
of the Pluto system, and for the satellite formation process in the Kuiper belt, 
are discussed in a companion paper\cite{Stern:2006}.

\end{boldmath}
\end{bfseries}



\begin{figure}
\bc
\includegraphics[angle=0,width=.45\textwidth]{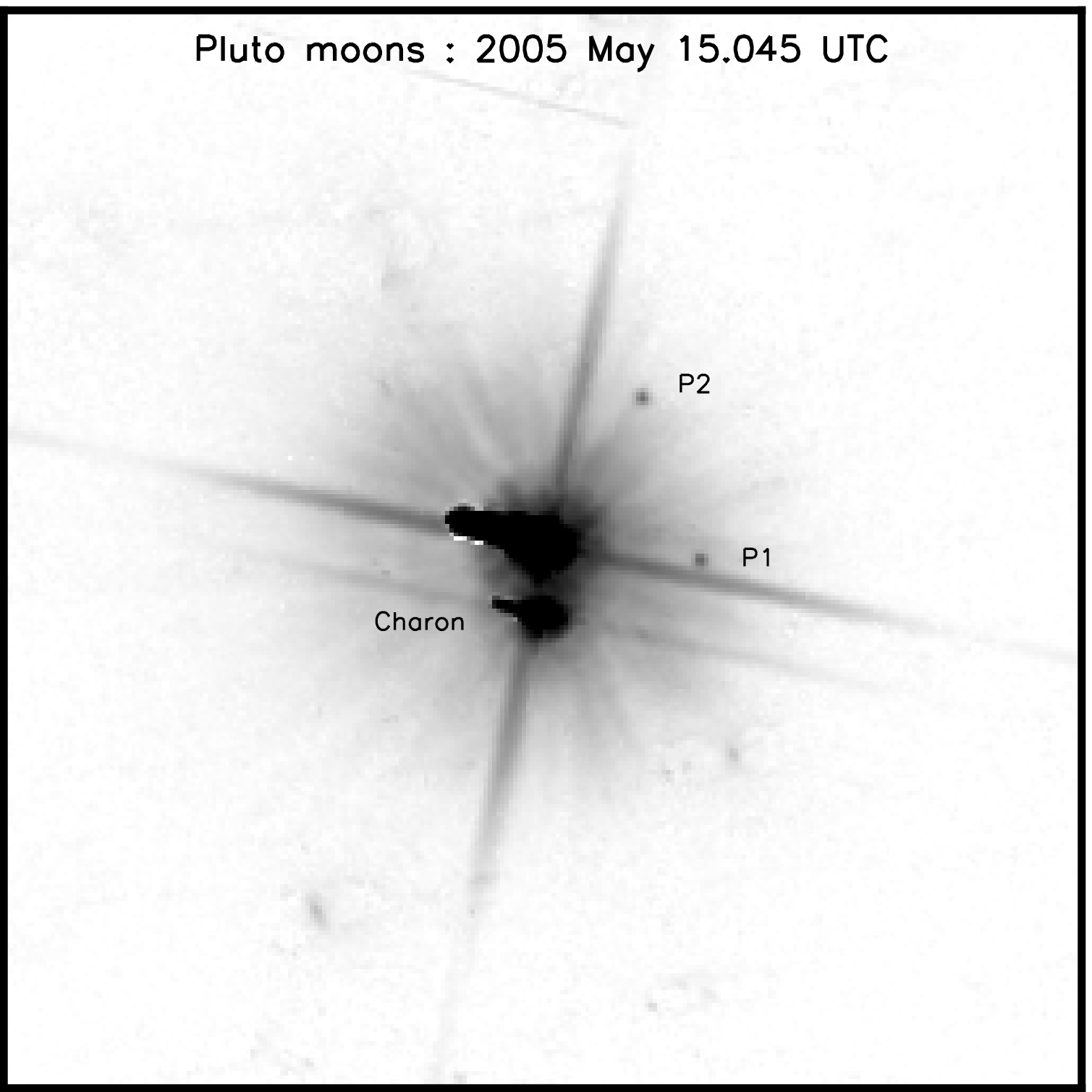}
\hfill
\includegraphics[angle=0,width=.45\textwidth]{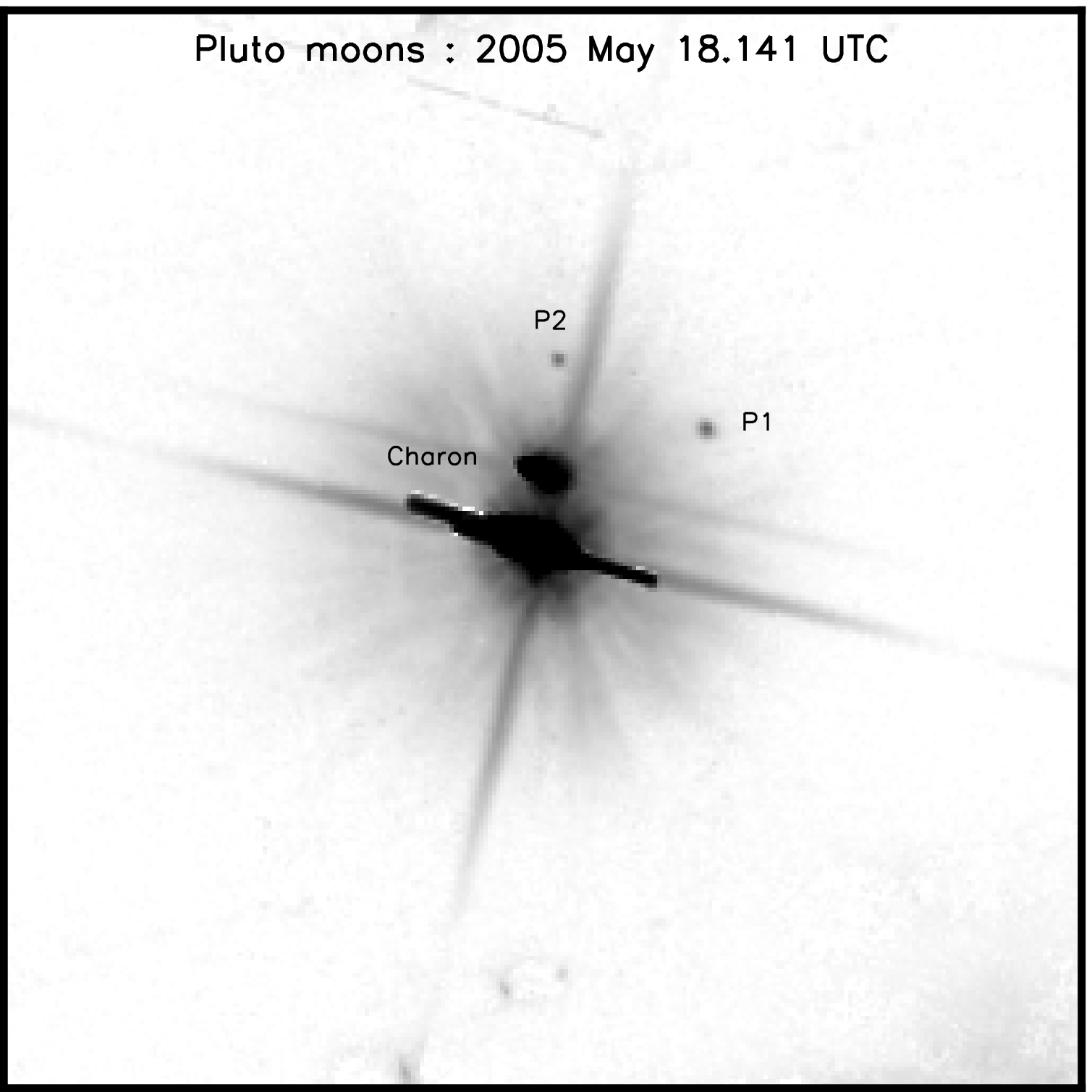}
\caption{
Images of the Pluto-Charon system showing the new satellites,
\mbox{S/2005 P 1} (P1) and  \mbox{S/2005 P 2} (P2).
The left frame is comprised of data taken on 2005 May~15.045 UT, and the right frame 
shows data taken 3.1 days later on 2005 May~18.141 UT.
Each frame is a small section \mbox{(256 pixel $\times$256 pixel)}
of the full image that is centered on Pluto and subtends
\mbox{12.8 arcsec $\times$ 12.8 arcsec}, which projects to
\mbox{280,000 km $\times$ 280,000 km} at Pluto.
Celestial north is up and east is to the left.
A logarithmic intensity stretch is used to enhance the visibility
of the two new satellites in the presence of the much brighter, saturated
images of Pluto and Charon.
P1 is near the 3~o'clock position on May~15 and moves to the 2~o'clock position 3.1 days later, while P2 moves from the 1~o'clock position to the 12~o'clock position over
the same time interval. Charon has moved (also counter-clockwise) from one side of Pluto to the other between the two epochs, as expected owing to its orbital period of 6.4~days. 
We used the individual, bias-subtracted, flat-fielded
files from the ACS calibration pipeline as the starting point for
our data reduction. We coaligned the four long-exposure images taken on each
date and combined them using a sigma-clipping procedure that filters out 
anomalous features
not present (to within the noise level) in all images. Thus, the many camera
artifacts present in the flat-fielded images (\eg cosmic ray events, hot pixels,
dead pixels, \ldots), as well as most of the star trails, are removed in the composite image. 
The few remaining faint blotches are artifacts caused by
incomplete subtraction of star-trails in the field, and their spatial distributions are
very different from the PSF-like distributions of P1 and P2.
The geometric distortion in the
composite image was removed using the same remapping algorithm employed 
by the standard ACS/WFC calibration process.
\label{fig:sats}}
\ec
\end{figure}

We observed Pluto with the \emph{Hubble Space Telescope (HST)}
using the Wide-Field Channel (WFC)
mode of the Advanced Camera for Surveys (ACS) on UT 2005 May~15 and May~18
(Fig.~\ref{fig:sats}).
The ACS/WFC consists of two \mbox{4096 $\times$ 2048 pixel} 
CCDs (WFC1 and WFC2) butted together,
effectively forming a single \mbox{4096 $\times$ 4096 pixel} camera with a gap
of $\sim$50~pixels between the two CCDs.
The F606W (``Broad V'') filter, which has a center wavelength of 591.8~nm and a width
of 67.2~nm, was used for all images.
At the time of the observations, Pluto was 31.0 astronomical units (AU) from the sun,
30.1~AU from the Earth, and had a solar phase angle of 0.96~deg on May~15
and 0.88~deg on May 18. 
Identical strategies were employed on each observing date. First, a single 
short exposure (0.5~s) was taken to enable accurate positions of Pluto and Charon 
to be measured on unsaturated images. Then, two identical, long
exposures (475~s) were taken at the same pointing to provide high sensitivity
to faint objects. Finally, the telescope was moved by $\sim$5~pixels in 
one dimension
and $\sim$60~pixels in the other dimension, and two identical, long exposures
(475~s) were taken to provide data in the region of the sky falling in
the inter-chip gap during the first two long exposures.
The telescope was programmed to track the apparent motion of Pluto 
\mbox{($\sim$3 arcsec hr$^{-1}$)} for all exposures. 

The two new satellites are detected with high signal-to-noise ratio
\mbox{($S/N \geq 35$)} and have a spatial morphology consistent with the
ACS point spread function (PSF; this is the spatial brightness distribution expected for 
unresolved objects).
Unlike virtually all of the spurious features in the individual ACS images,
the two new objects were observed in two consecutive images on two
different dates.
Furthermore, the objects do {\it not} appear in the images where they are not expected,
namely when the telescope pointing placed the objects in the inter-chip gap.
Although P1 and P2 are several thousand times fainter than Pluto,
their images look nothing like the well-documented\cite{Hartig:2003}
ghosts and scattered-light artifacts produced near
bright, highly saturated objects.
In short, all the available evidence 
supports the claim that the objects observed near Pluto-Charon are real 
astronomical objects, and none of the data suggest that these objects are
observational artifacts.

Next we address the issue of whether P1 and P2 are associated with the Pluto-Charon
system.
Only astronomical bodies moving at approximately the same non-sidereal rate as Pluto
will have PSF-like spatial distributions, and asteroids and stars are easily recognized
by their trailed images. 
We estimate that foreground or background objects must have a heliocentric
distance within  $\sim$0.25~AU of Pluto's to remain within 2~arcsec
of Pluto over the 3.1~days between the two sets of ACS observations.
Based on the known sky-projected density of KBOs as a function of 
magnitude\cite{Bernstein:2004}, the
probability of finding {\it any} similarly bright KBO 
within 2.5~arcsec of  Pluto is \mbox{$\sim$$4 \times 10^{-6}$,} which means
that the probability of finding {\it two} KBOs in this same region is 
\mbox{$\sim$$1.6 \times 10^{-11}$.}
These statistical arguments are not strictly applicable to the plutino KBO population
(\ie objects that share Pluto's 3:2 orbital resonance with Neptune),
for which resonances and other gravitational perturbations might keep objects
relatively close to Pluto. Nevertheless, there must be only a vanishingly small chance 
that the two observed objects are random 
KBOs that happen to be aligned with Pluto during the course of our observations.

\begin{figure}
\bc
\includegraphics[angle=0,width=.6\textwidth]{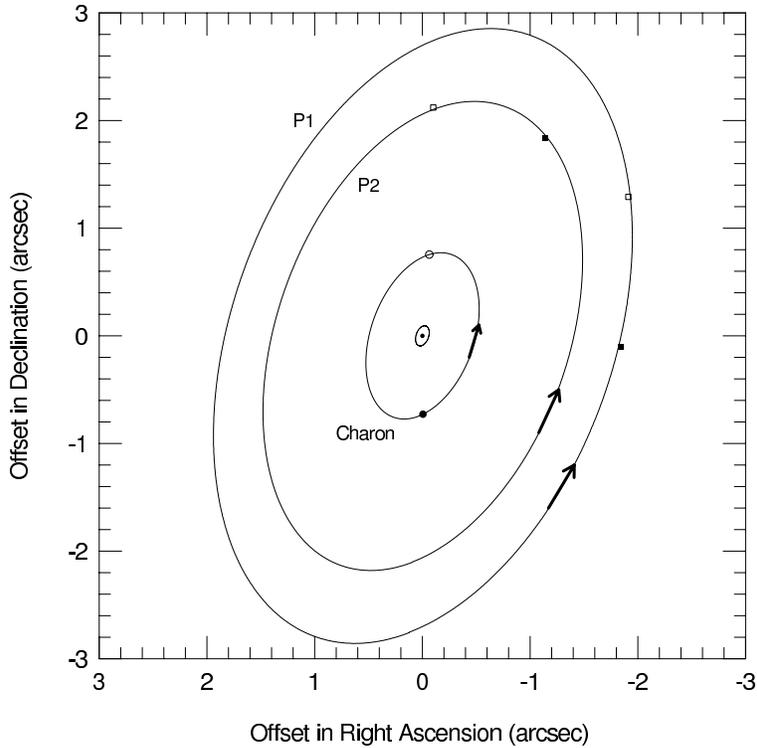}
\caption{
Preliminary orbits for P1 and P2, assuming they are circular and in the same plane as Charon's orbit.
The barycenter of the system is the dot in the center, Pluto's orbit is the smallest ellipse, Charon's orbit is the next ellipse (its positions on May 15 and May 18 are indicated by the filled and open circles, respectively), an orbit that is consistent with P2's measured positions is next, followed by an orbit that is consistent with P1's measured positions. 
For both of the latter cases, the filled squares are the observed positions on May 15, and the open squares are the observed positions on May 18.
The uncertainties (1$\sigma$) in the measured positions are approximately the 
size of the symbols.
At Pluto's distance, 1~arcsec projects to 21,800~km. Charon's orbital plane is inclined by 
$\sim$36~deg relative to the line-of-sight, which means that the circular orbits of the satellites appear elliptical when projected on the plane of the sky.
All four objects orbit the barycenter in a counter-clockwise direction in this view,
and all positions are displayed in the J2000 reference frame with celestial north up
and east to the left.
\label{fig:orbits}}
\ec
\end{figure}

Orbital analysis based on the astrometry of P1 and P2 (Table~\ref{tbl:photometry}) provides even stronger evidence that the two objects are indeed associated with Pluto.
Although the astrometric data from two observations alone are not enough to compute definitive orbits, we investigated the hypothesis that the objects have circular 
orbits in the same plane as Charon's orbit. Adding these two constraints
(\ie zero eccentricity and the same orbital plane as Charon) allows us to solve 
for the orbital radii that best match the observations. 
We find that the orbital radii are
\mbox{64,700 $\pm$ 850 km} for P1 and 
\mbox{49,400 $\pm$ 600 km} for P2, both measured relative to the system barycenter.
The results are depicted in
Fig.~\ref{fig:orbits}. The excellent agreement between the observed and model
orbits \mbox{(reduced $\chi_{\nu}\!^{2} \approx 1$} for the fits)
supports our assumption that the orbits are 
circular and coplanar with Charon's orbit, and also solidifies the conclusion
that P1 and P2 are associated with the Pluto system.
Using the known mass of the Pluto system and Kepler's third law,
we derive orbital periods of
\mbox{38.2 $\pm$ 0.8 days} for P1 and \mbox{25.5 $\pm$ 0.5 days} for P2.
The ratios of the orbital periods to the orbital period of Charon 
\mbox{(6.387245 $\pm$ 0.000012 days)\cite{Buie:1992}} are
\mbox{5.98 $\pm$ 0.12} for P1 and \mbox{3.99 $\pm$ 0.07} for P2. This suggests that
P1 may be in a 6:1, and P2 may be in a 4:1, mean motion orbital resonance with Charon, and that P1 and P2 may be in a 3:2 resonance with each other. 
Further investigation of these possible resonances, or slight deviations from these
resonances, will undoubtedly provide significant insights into the 
dynamical evolution of the Pluto system.

After finding the orbital solutions described above, we re-examined some \hst data of
Pluto taken in 2002 under another program (Buie, Principal Investigator). 
That program was optimized to study
the surfaces of Pluto and Charon and used much shorter exposure times (6~sec for
a V-band filter and 12~sec for a B-band filter) than employed in our program.
Nevertheless, by combining all the images obtained on a given date in the
same filter, a limiting sensitivity could be achieved that is very close to 
the observed magnitudes of
P1 and P2 (see Table~\ref{tbl:photometry}). On the observation date 
\mbox{(UT 2002 June 14)} with the
best geometry (lowest phase angle and smallest geocentric distance), two objects
are weakly detected (S/N$\approx$4) along the predicted orbital paths of P1
and P2 in both the V-band and B-band images, providing independent evidence
that P1 and P2 are indeed satellites of Pluto.


Our photometry results are summarized in Table~\ref{tbl:photometry}.
On May~15 P2 has a visual magnitude ($V$) of \mbox{$23.38 \pm 0.17$, } 
and on May~18 P1 
has \mbox{$V = 22.93 \pm 0.12$.}
P1 is too close to a diffraction spike for accurate photometry on May~15, but the
brightness then seems consistent with the value observed on May~18.
P2 is too close to a diffraction spike for accurate photometry on May~18, but the brightness then seems consistent with the value observed on May~15. Small bodies are often highly
irregular in shape, which results in significant temporal variation in brightness as the object rotates. Whether this is true for P1 and P2 must await future investigations.

If the geometric albedo is known, the size of an object can be calculated from its magnitude
using a standard relation\cite{Russell:1916}.
For a geometric albedo of 0.04 (\ie comet-like\cite{Lamy:2004}), P1
has a diameter of \mbox{167 $\pm$ 10 km,} and P2 has a diameter
of \mbox{137 $\pm$ 11 km}. 
If the albedo is 0.35 (\ie Charon-like\cite{Marcialis:1992}), the diameters are 
\mbox{61 $\pm$ 4 km} for P1 and \mbox{46 $\pm$ 4 km} for P2.
Both satellites are tiny compared to Pluto and Charon, which have diameters of 
\mbox{2328 $\pm$ 42 km\cite{Buie:1992,Young:1994a}}
and
\mbox{1208 $\pm$ 4 km\cite{Sicardy:2005},} respectively.
The masses of P1 and P2 account for less than
\mbox{$5 \times 10^{-4}$}
of the mass of the Pluto system, assuming that the densities 
of P1 and P2 are similar to, or smaller than, Pluto's density.

We also searched the entire ACS/WFC field-of-view
\mbox{(202 arcsec $\times$ 202 arcsec)} for other satellites,
but none were found down to a limiting magnitude of
\mbox{$V \leq 26.2$} (90\% confidence limit) for
the region between 5 and 100 arcsec from Pluto\cite{Steffl:2005}.
Thus, the Pluto system of satellites appears to be very compact.


\vspace*{1in}

\bibliography{bibtex_refs}


\noindent
{\bf Acknowledgments}
We thank George Hartig for extensive discussions of the ACS optical performance
and for examining the images discussed here.
We thank the Directors and
staff at the Keck, Very Large Telescope, and Gemini observatories for their
heroic efforts in attempting ground-based recoveries
of these new satellites under non-optimal conditions.
We thank the Director and staff of the STScI for their excellent support of 
the \hst observations.
Support for this work was provided by NASA through a grant
from the Space Telescope Science Institute, which is operated by the
Association of Universities for Research in Astronomy, Inc., under NASA
contract.\\

\noindent
{\bf Author Information}
Reprints and permissions information is available at
npg.nature.com/reprintsandpermissions. The authors declare no competing
financial interests. Correspondence and requests for materials should be
addressed to H.A.W.~(email: hal.weaver@jhuapl.edu).

\clearpage

\begin{deluxetable}{ccccccccc}
\tablecaption{Astrometry and Photometry of Pluto Satellites.
\label{tbl:photometry}
}
\tablecolumns{9}
\tabletypesize{\footnotesize}
\tablewidth{0pc}
\rotate
\tablehead{
\colhead{\bf Date}
	& \multicolumn{2}{c}{\bf Pluto}
	& \multicolumn{2}{c}{\bf Charon}
	& \multicolumn{2}{c}{\bf S/2005 P 1}
	& \multicolumn{2}{c}{\bf  S/2005 P 2}\\

\colhead{UT May 2005}
	& \colhead{[$r$,$\theta$]}		& \colhead{$V$}
	& \colhead{[$r$,$\theta$]}		& \colhead{$V$}
	& \colhead{[$r$,$\theta$]}		& \colhead{$V$}
	& \colhead{[$r$,$\theta$]}		& \colhead{$V$}
}

\startdata

15.045	
	& [0,0]					& 14.12 $\pm$ 0.08
	& [0.834,176.77]			& 16.22 $\pm$ 0.10
	& [1.85,264.21]				& \nodata
	& [2.09,326.94]				& 23.38 $\pm$ 0.17\\
	
18.141	
	& [0,0]					& 14.28 $\pm$ 0.08
	& [0.855,355.04]			& 16.28 $\pm$ 0.10
	& [2.36,305.76]				& 22.93 $\pm$ 0.12
	& [2.22,355.51]				& \nodata

\enddata

\tablecomments{
``Date'' refers to the mid-point of the observing period. $r$ is the distance
in arcsec and $\theta$ is the celestial position angle in degrees (measured eastward 
from north) in the J2000 reference frame from the center-of-light of the specified 
object to the center-of-light of Pluto.
The 1$\sigma$ errors in $r$ and $\theta$ are 0.035~arcsec and 0.7~deg, respectively,
and are mainly due to the uncertainty in locating Pluto in the long-duration exposures.
The center-of-light is offset from the center-of-body by $\sim$100~km for Pluto
and by $\leq$20~km for Charon\cite{Buie:1992}, but these offsets were ignored in
our orbital analyses.
$V$ is the observed visual magnitude of the object, when available, and it was
calculated using a 5-pixel-radius (0.25~arcsec) aperture.
We used the information and procedures described in Sirianni \etal\cite{Sirianni:2005}
to convert the observed signal in this small aperture to a V-magnitude in the
standard Johnson-Landolt system. 
Adopting a \mbox{$B$$-$$V$} color of 0.7
(\ie similar to Charon-like\cite{Marcialis:1992} and solar-like\cite{Allen:1976} colors), 
we derive
\mbox{$V = -2.5 \log S/t + (26.38 \pm 0.07)$,}
where $V$ is the V-band magnitude in the Johnson-Landolt system, $S$ is the net signal in
electrons in the 5-pixel-radius aperture, and $t$ is the exposure time in seconds.
We calculated the magnitudes of Pluto and Charon the same way, although
both objects (especially Pluto) are slightly resolved, 
which introduces an additional systematic error in those cases.
}
\end{deluxetable}

\clearpage

\clearpage

\end{document}